# Environmental, Thermal, and Electrical Susceptibility of Black Phosphorus Field Effect Transistors

Zenghui Wang, Arnob Islam, Rui Yang, Xu-Qian Zheng, Philip X.-L. Feng[*]

*Department of Electrical Engineering & Computer Science, Case School of Engineering,*

*Case Western Reserve University, 10900 Euclid Avenue, Cleveland, OH 44106, USA*

**Atomic layers of black phosphorus (P) isolated from its layered bulk make a new two-dimensional (2D) semiconducting crystal with sizable direct bandgap, high carrier mobility, and promises for 2D electronics and optoelectronics. However, the integrity of black P crystal could be susceptible to a number of environmental variables and processes, resulting in degradation in device performance even before the device optical image suggests so. Here, we perform a systematic study of the environmental effects on black P electronic devices through continued measurements over a month under a number of controlled conditions, including ambient light, air, and humidity, and identify evolution of device performance under each condition. We further examine effects of thermal and electrical treatments on inducing morphology and, performance changes and failure modes in black P devices. The results suggest that procedures well established for nanodevices in other 2D materials may not directly apply to black P devices, and improved procedures need to be devised to attain stable device operation.**

## I. Introduction

Two-dimensional (2D) crystals isolated from their layered bulk materials, and their heterostructures via 'lego-stacking' assembly, hold great promises for future electronic and optoelectronic devices/transducers, thanks to the many attractive physical properties and ultimate scalability set by the discrete atomic layers. For example, graphene, the pioneering 2D material, exhibits ultrahigh carrier mobility (~$10^5$ cm²V⁻¹s⁻¹) even at room temperature,[1,2] desirable for high-speed electronics. However, the lack of bandgap (plus challenges in practical ways of creating a usable bandgap) limits its application in logic electronics. The later introduced 2D transition metal di-chalcogenides (TMDC),[3,4] with representative members including molybdenum disulfide (MoS₂) and tungsten diselenide (WSe₂), boast sizable bandgaps and have facilitated impressive and encouraging progresses in 2D logic circuits.[5,6] Nonetheless, the carrier mobilities in 2D TMDCs (typically ~$10^2$ cm²V⁻¹s⁻¹) are still comparatively much lower than those of graphene, and remedies are desirable to boost their performance toward 2D radio frequency (RF) circuits and high-speed logic circuits.[7]

---

[*]Corresponding Author. Email: philip.feng@case.edu

Black phosphorus (P) atomic layers has recently been isolated from its layered bulk and emerged as a new 2D single-element semiconductor.[8] It is found to possess a number of notable advantages over other existing 2D materials. In particular, black P has a layer-number-tunable direct bandgap (the optical bandgap ranges from ~1.5eV in monolayer to 0.3eV in bulk) that covers visible to infrared,[9,10,11,12,13,14,15,16,17,18] and simultaneously high carrier mobility (up to 4000 $cm^2V^{-1}s^{-1}$, which has enabled observation of quantum oscillations[19,20]) with expected values approaching that of graphene.[21] However, recent studies indicate that ultrathin black phosphorus may not be stable in ambient conditions, which casts a shadow upon developing processes toward reliable and robust devices.[22] Optical characterization of black P suggests that the coexistence of light, air, and moisture can lead to degradation in the phosphorus crystal, while exposure to each individual element might leave the layered material intact.[23] Here, we describe systematic characterization of black P field effect transistor (FET) devices for their susceptibility to visible light, air, and moisture, through careful control of the residing environment of the device and continued measurements spanning more than a month, for multiple devices with and without protective covers. We further investigate the effects on the morphology and performance of black P FET devices from thermal annealing and application of gate polarization, which are processes common to 2D device fabrication and characterization. Our findings suggest that black P FET devices are highly susceptible to environmental disturbances, and demand sustained precaution throughout device fabrication, measurement, and storage.

## II. Experimental Results and Discussions

### A. Device Fabrication

We fabricate black P FETs with and without hexagonal boron nitride (h-BN) protective cover layer by using a dry transfer method,[24] with a representative set of devices shown in Fig. 1. First, we pattern the electrodes on a 290nm- $SiO_2$-on-Si substrate by using photolithography followed by metallization (5nm Cr/30nm Au) and liftoff, at wafer scale. Some substrates are further patterned with photolithography followed by reactive ion etching (RIE) to create microtrenches (190nm deep) in the $SiO_2$ layer between the electrodes for making suspended devices. Black P flakes and h-BN flakes are exfoliated onto polydimethylsiloxane (PDMS) stamps[24,25] and are identified using optical microscope. Potential candidates are chosen based on flake thickness (inferred from color under microscope), size, and shape. The receiving substrate is then chosen based on the geometry of the 2D flakes. Using a micropositioner the black P flake on PDMS stamp is first aligned to and then transferred onto the electrodes to make FET devices (Fig. 2a). Some devices undergo a second alignment and transfer of an h-BN protective cover layer, resulting in h-BN-covered black P FETs (Fig. 2b). By choosing the proper substrate and 2D flakes, it is possible to fabricate multiple devices through a single set of dry transfer procedures. The transfer processes shown in Fig. 2 results in the devices shown in Fig. 1 with multiple black P FETs connected by a group of electrodes, and h-BN covering roughly half of the black P devices. From atomic force microscopy (AFM) measurements we determine that the black P is 12nm in thickness, and the h-BN protective cover is 15nm thick.

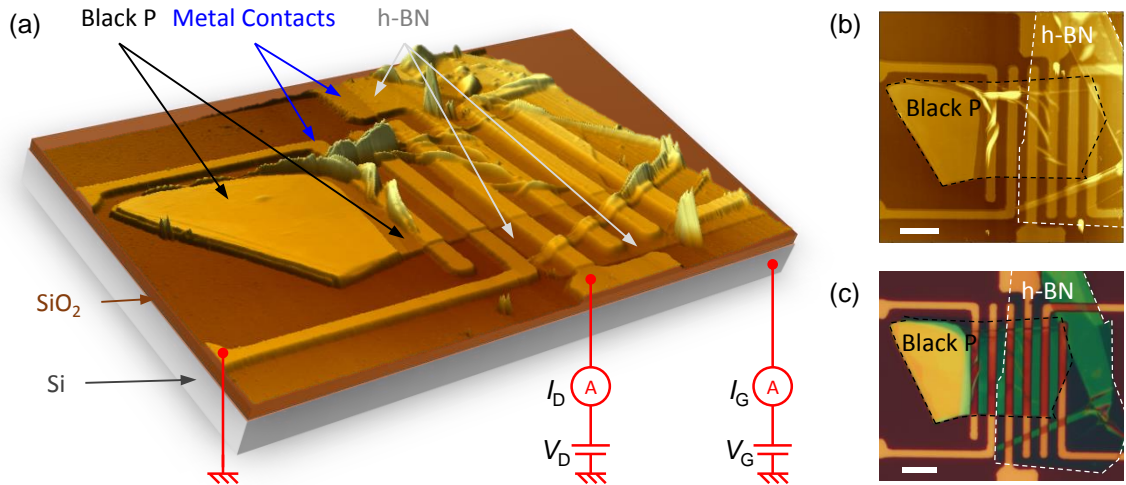

**Figure 1.** Black phosphorus (P) field effect transistor (FET) devices and schematic of electronic characterization. (a) 3D AFM image of a group of black P FETs, with an h-BN protective layer covering several of the devices. Schematic electrical connection during a typical measurement is shown for one FET device. (b) Top-view AFM image and (c) optical image of the same set of devices.

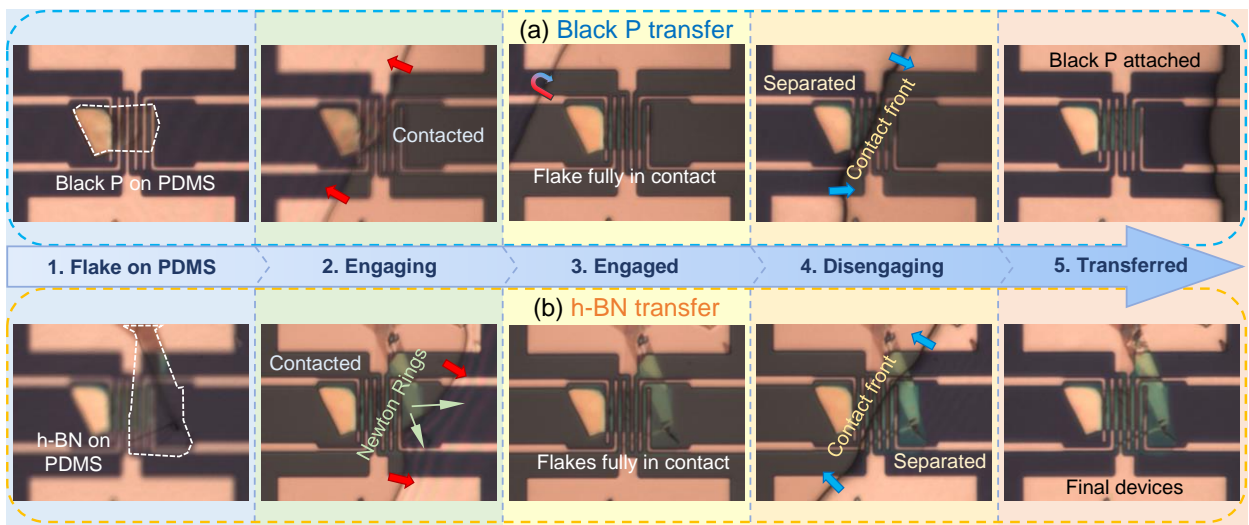

**Figure 2.** Steps of dry transfer process for making black P FET devices. (a) Transferring black P flake onto a set of electrodes. (b) Transferring h-BN to cover part of the black P flake.

## B. Electronic Characterization and Environmental Control

To evaluate the effect of different environmental variables on the electronic performance of black P FETs, we subject the devices to a series of environmental configurations, and monitor their electronic performance through continued measurements.

We control exposure to light by storing devices in black plastic boxes (MK-51 Mini Box, Electronics USA) (Fig. 3a). The exposure to ambient air is removed by storing sample inside a vacuum desiccator (Bel-Art Scienceware Polycarbonate Techni-Dome) (Fig. 3b) connected to a vacuum pump. The ambient air in the lab typically has a humidity of 20%. During air exposure, the humidity can be boosted by placing the devices ~1 foot downstream of a cool-mist humidifier (HoMedics HUM-CM10) (Fig. 3c). We use a probe station connected to a Keithley 4200 semiconductor characterization system for electronic measurements. To minimize exposure to ambient light, the probe station door is closed to keep the device in darkness during measurements (Fig. 3d&e).

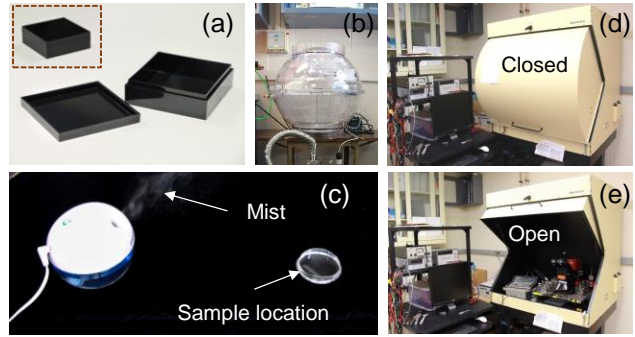

**Figure 3**. Environmental controls. (a) Black storage box with lid used to protect samples from visible light. *Inset*: Closed box. (b) Vacuum desiccator used to store sample in vacuum. (c) Humidifier used to increase sample exposure to humidity. (d) & (e) The probe station and semiconductor parameter analyzer system used for electrical measurements.

Figure 4a shows the environmental contingency experienced by the group of devices (corresponding to the different sections labeled in Fig. 4a): (I) upon fabrication, it is first kept in vacuum and darkness for 11 days to establish a baseline; (II) the sample is then moved out of the vacuum desiccator for air exposure, while still kept in darkness; (III) after 68 hours of air exposure, it is moved back into vacuum and inside black box to obtain the $2^{nd}$ baseline; (IV) after another 6 days, the black box is removed, and the black P FETs are exposed to room light while being kept in vacuum; (V) then another baseline (in vacuum, no light) of 6 days is measured, before (VI) finally the devices are completely exposed to both light and ambient air. During this final exposure, periodic humidity increases are introduced using the humidifier.

Intermittent electrical measurements (with typical connection shown in Fig. 1a) are taken during the baselines and the different exposure phases, until the electronic performance of the devices degrades completely (*i.e.*, with device electrical conductivity falling below measurement sensitivity). Throughout the experiment we measure two types of electronic responses of the black P FETs: (i) transport characteristics (Fig. 4(b)), where we sweep the drain bias $V_D$ and measure the drain current $I_D$ under a series of gate voltage $V_G$, and (ii) transfer characteristics (Fig. 4(c)), where we measure $I_D$ as a function of $V_G$ under a given $V_D$ (we sometimes adjust $V_D$ to accommodate the difference between individual devices as well as evolution in each device's characteristics).

### C. Characterization of Degradation under Varied Conditions

To quantify the device performance in order to evaluate the effect of environmental variables on black P FETs, we parameterize the transport characteristics with 6 derived values from the measurements: (**1**). $R$, the small-bias resistance at $V_G$=-10V (the low-bias slope of the corresponding transport curve, dashed line in Fig. 4(d) left pane); (**2**). $I_D$ at the maximum forward bias ($V_{D,max}$) at $V_G$=-10V (circled data point in Fig. 4(e) left pane); (**3**). $I_D$ at the maximum reverse bias (negative $V_{D,max}$) at $V_G$=-10V (circled data point in Fig. 4(f) left pane);

**(4)**. $I_D$ at $V_G$=-10V in the $V_G$ sweep (on the retrace of the sweep when $V_G$ decreases from 10V to -10V), which is typically the highest current value during the $V_G$ sweep (circled data point in Fig. 4(g) left pane); **(5)**. $I_D$ at $V_G$=10V in the $V_G$ sweep, which is typically the lowest current value during the $V_G$ sweep (circled data point in Fig. 4(h) left pane); **(6)**. $\Delta I_D$, the difference between the $I_D$ values at $V_G$=-10V and $V_G$=10V, which represents the gate tunability in this $V_G$ range (as indicated by the double-ended arrow in Fig. 4(i)). We then plot these extracted parameters as functions of time with each section (environmental condition) labeled (Fig. 4 (d)-(i) center panes), and expand the last section (with both air and light exposure) into zoomed-in plots (Fig. 4 (d)-(i) right panes) to show the details of the final degradation (VI).

Figure 5 shows the parameterized result summaries for three black P FETs fully covered by the h-BN protective layer, and the results for uncovered (or partially covered) devices undergone the same environmental contingency are shown in Fig. 6. From the data we make the following observations, and discuss each of them below.

First, both h-BN covered and uncovered black P FET devices eventually degrade under extended exposure to air, light, and humidity, with $R$ values significantly increasing and with $I_D$ and $\Delta I_D$ values diminishing. While the h-BN layer does not protect the black P from eventual degradation, the data does suggest that noticeable deterioration of electronic transport properties (*e.g.*, increase in $R$, decrease in $I_D$) takes place earlier in uncovered devices (starting around 10~30 hours in the right panes of Fig. 6) than in fully covered devices (no observable degradation until ~70 hours). This suggests that even though the h-BN protective cover used in this device geometry (electrodes at bottom) may not completely seal the black P flake (with possible leak paths at the edge of electrodes where the 2D materials might not fully conform to the local detailed shapes or terrains of the electrodes), the h-BN cover does to some extent impede the degradation of black P through contact with the environment. Therefore, using different device geometries, such as fully-encapsulated structures, should further improve the stability of black P FET devices.

Second, exposure to air can lead to degradation in the electronic performance of black P FETs even in the absence of visible light (section II in Fig. 5 & Fig. 6). Such effects are not caused by the intermittent electrical characterizations, as the same type of measurements in other sections (environmental configurations) show no sign of degradation. Earlier experiment has found no change in the $A_g^2$ Raman peak intensity in black P flake under similar conditions on the order of hours.[23] Our results show that while the presence of light can accelerate the degradation process, the reaction between black P and the ambient environment can still take place, though much slower, even without the presence of light. In contrast, when black P FET is exposed to light illumination but not the ambient (section IV in Fig. 5 & Fig. 6), no degradation in electronic transport is observed. This is consistent with the fact that the chemical reactions cannot proceed without the reactants—oxygen and water vapor—oxygen and water vapor, both of which are known to react with black P samples (producing $P_xO_y$ species as verified by electron energy loss spectroscopy[23])—even when sufficient energy would be provided by the photons. Additionally, surface adsorption can lead to change in device performance,[24] which again does not require light illumination.

Third, the appearance of the device remains largely unchanged under optical microscope despite measurable degradation in device performance. Figure 7 shows the optical images of the device taken throughout the entire measurement. No noticeable change is observed, until near the end towards complete degradation (note that the small water droplets in the last few images are from the cool-mist humidifier, which are of the size ~1μm, and are evenly distributed over the electrodes, substrate, and the device surface). We note that a number of physical and chemical processes can be involved in the degradation (physical adsorption, chemical adsorption, surface wetting, oxidation, *etc.*), and some processes may have greater effect on device appearance while others on device performance, thus resulting in the lack of apparent correlation between the two. Our measurement shows that degradation in device performance can take place long before changes can be noticed in device appearance, and thus precaution must be taken in monitoring and evaluating the degradation of black P devices, not just by optical microscope inspections.

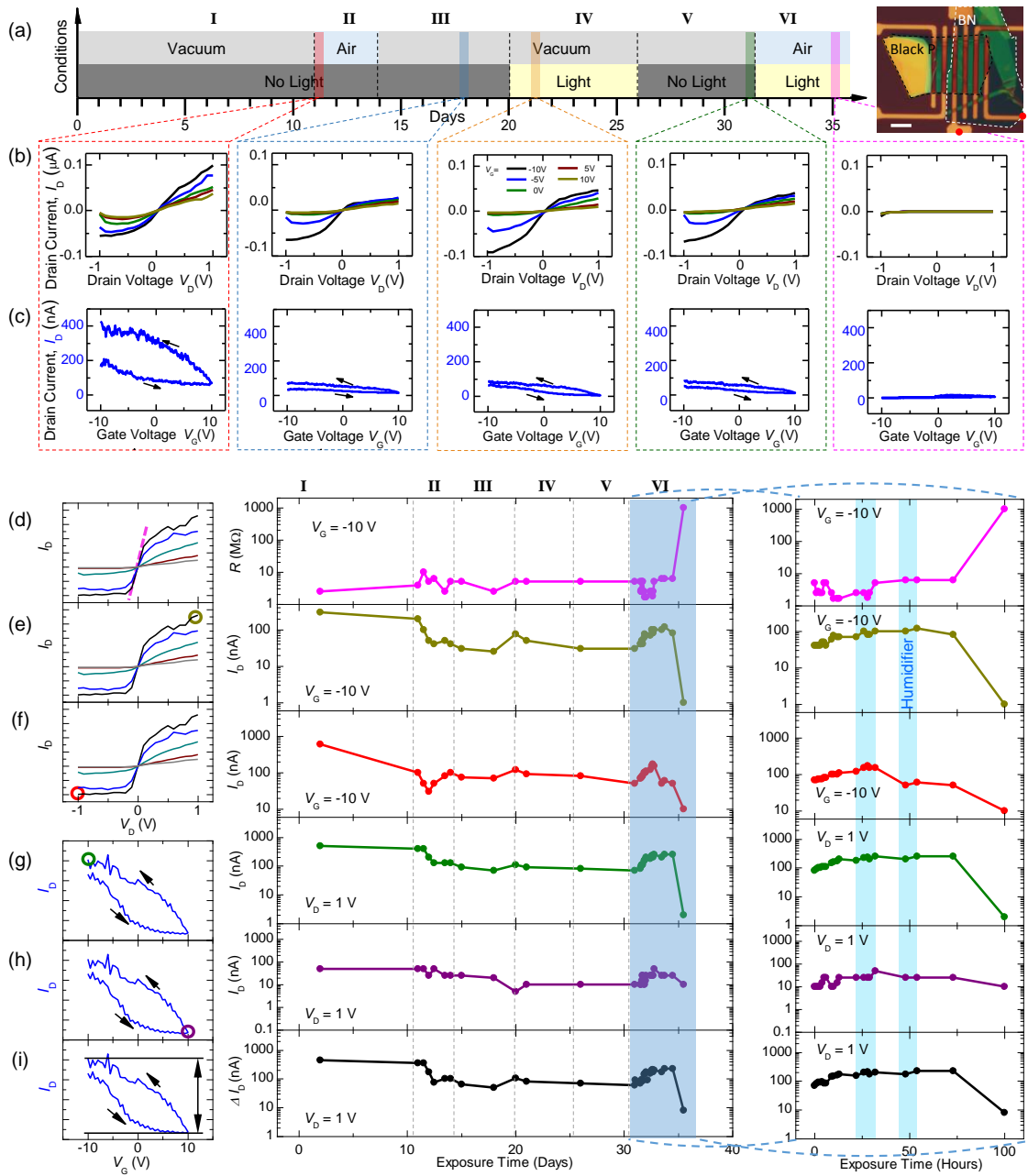

**Figure 4.** Electrical measurement of black P FET device degradation. (a) Evolution of environmental variables over time, with I-VI denoting the six different environmental contingencies. The data in (b)-(i) are measured between the pair of electrodes as indicated by the red dots in the image (scale bar: 10µm). Example (b) transport curves and (c) transfer curves are shown for different stages during the measurement, from which we extract a number of parameters: (d) small-bias resistance at $V_G$=-10V; current at largest (e) forward and (f) backward bias at $V_G$=-10V during $V_D$ sweeps; current at (g) $V_G$=-10V (from retrace) and (h) $V_G$=+10V during $V_G$ sweeps, and (i) their difference. The left panels illustrate how the parameters are extracted from respective measurements, and the center panel plots each parameter as a function of time over the entire course of measurement, with a zoomed-in view of the final degradation stage (VI) shown in the right panels.

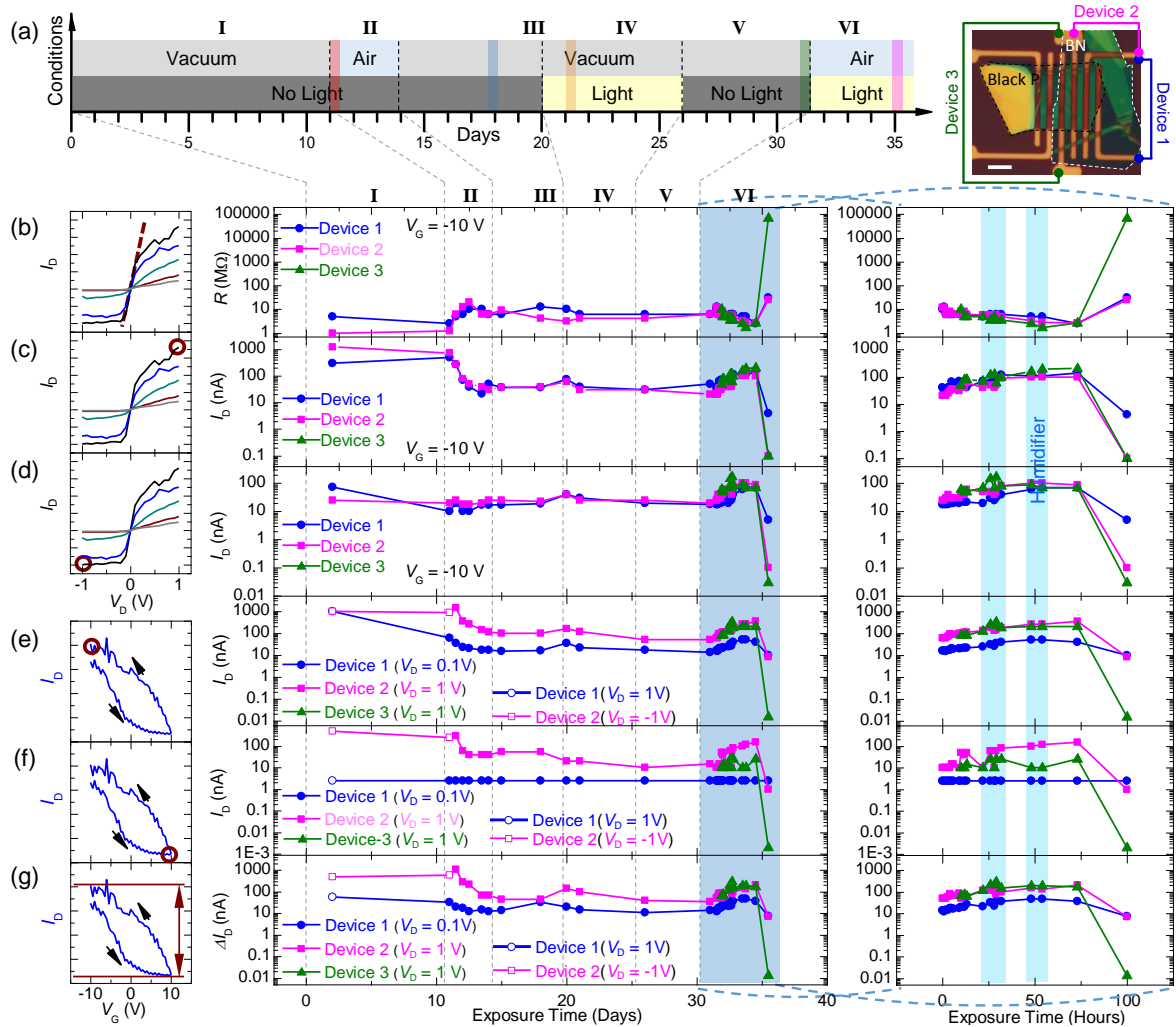

**Figure 5.** Degradation of h-BN covered black P FETs. (a) Evolution of environmental variables over time, with I-VI denoting the six different environmental contingencies. The optical image shows the different h-BN covered FET devices measured (scale bar: 10µm). The plotted parameters are: (b) small-bias resistance at $V_G$=-10V; current at largest (c) forward and (d) backward bias at $V_G$=-10V during $V_D$ sweeps; current at (e) $V_G$=-10V (from retrace) and (f) $V_G$=+10V during $V_G$ sweeps, and (g) their difference. The left panels illustrate how the parameters are extracted from respective measurements (for illustration purpose only; do not correspond to a specific device or measurement), and the center panel plots each parameter as a function of time over the entire course of measurement, with a zoomed-in view of the final degradation stage (VI) shown in the right panels.

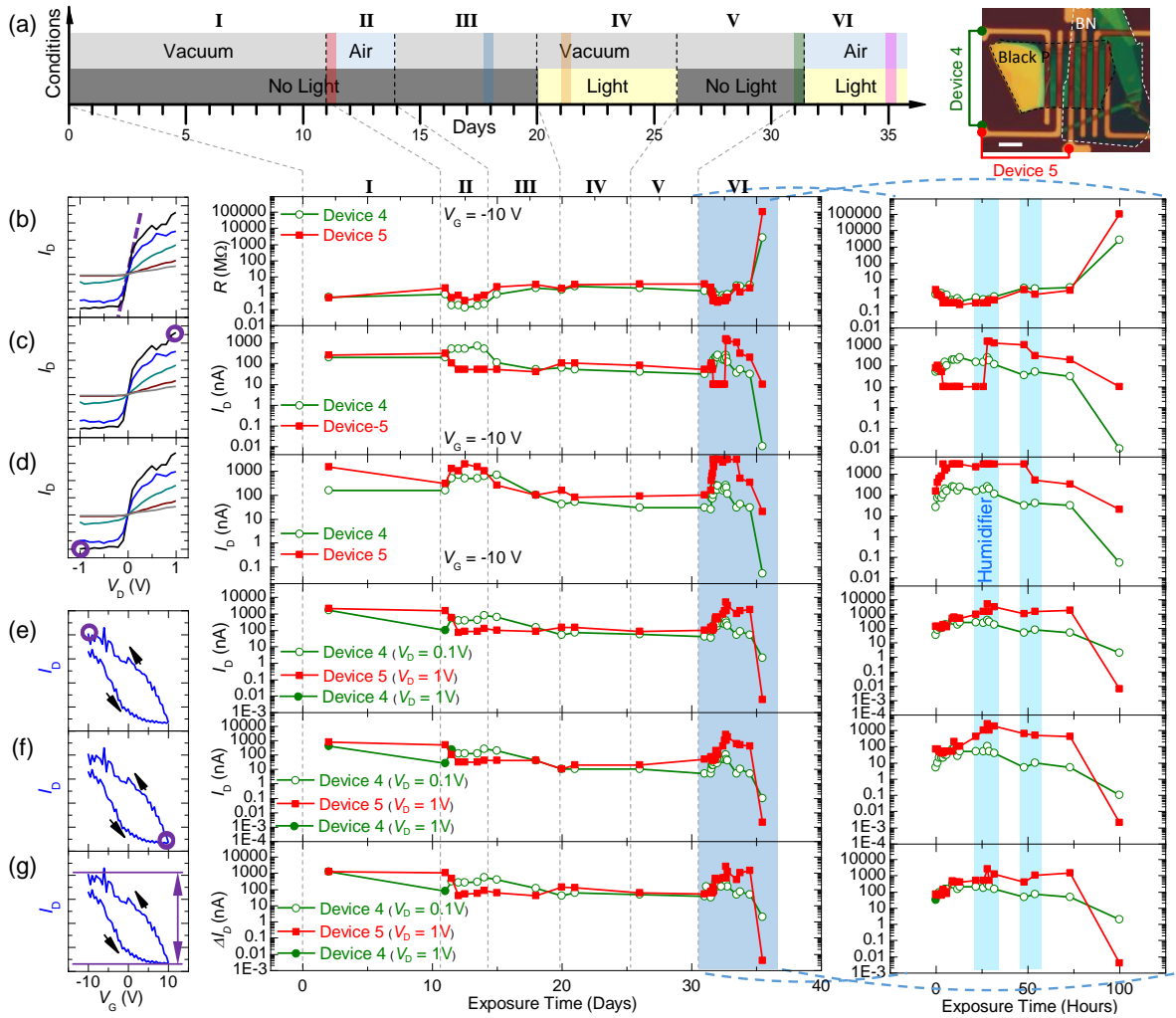

**Figure 6.** Degradation of uncovered black P FETs. (a) Evolution of environmental variables over time, with I-VI denoting the six different environmental contingencies. The optical image shows the different uncovered FET devices measured (scale bar: 10µm). The plotted parameters are: (b) small-bias resistance at $V_G$=-10V; current at largest (c) forward and (d) backward bias at $V_G$=-10V during $V_D$ sweeps; current at (e) $V_G$=-10V (from retrace) and (f) $V_G$=+10V during $V_G$ sweeps, and (g) their difference. The left panels illustrate how the parameters are extracted from respective measurements (for illustration purpose only; do not correspond to a specific device or measurement), and the center panel plots each parameter as a function of time over the entire course of measurement, with a zoomed-in view of the final degradation stage (VI) shown in the right panels.

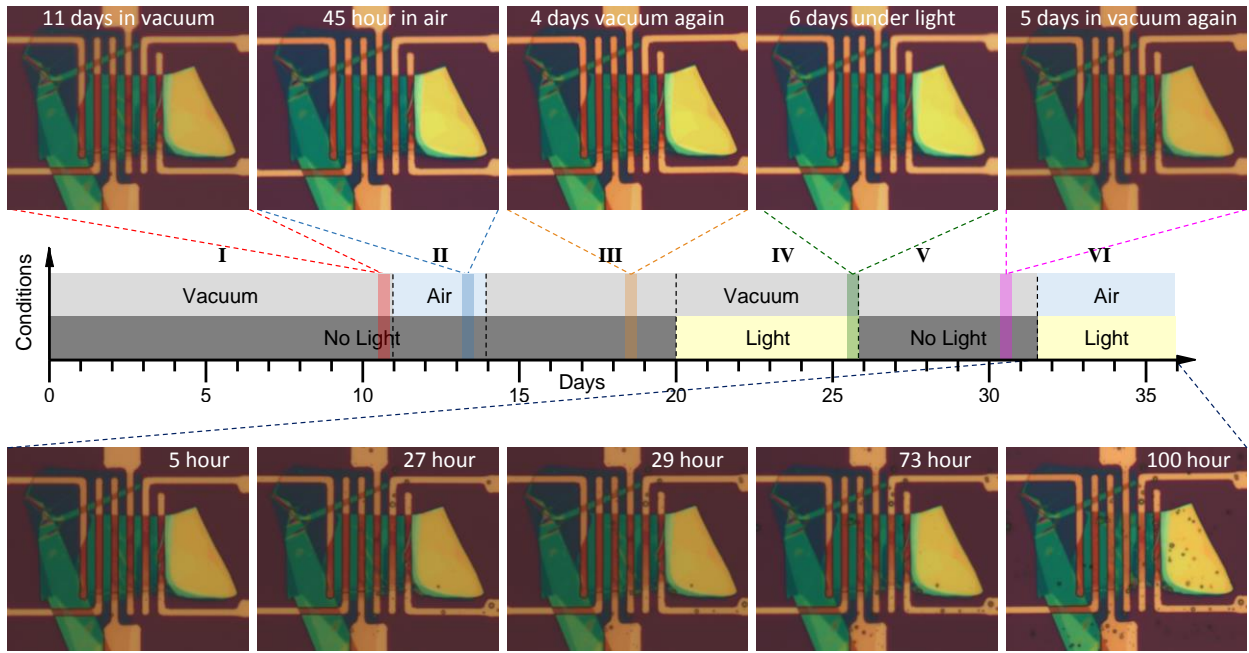

**Figure. 7.** Optical images of the devices shown in Fig. 4-6 throughout the measurement. The middle panel shows the evolution of environmental variables over time, with I-VI denoting the six different environmental contingencies. The top panel shows example images during stages I-V, and the bottom panel shows example images during the final degradation stage (VI). Note that humidifier is used intermittently during the final stage, resulting in water droplets on the substrate and electrodes.

## D. Investigation of Thermal Annealing Effects

We now turn to the effects from thermal annealing on black P FET devices. Annealing is a common practice in 2D device fabrication and treatment (*e.g.*, often used in cleaning and restoring graphene, $MoS_2$ and other 2D devices), and is shown to enhance device performance in 2D TMDC FETs fabricated using the same dry transfer method by improving contact quality and removing adsorbates.[24] However, given the different chemical properties between black P and TMDC, the effects of thermal annealing on these 2D materials can be different, and need to be characterized respectively.

Figure 8(a) shows a group of black P devices, several of which are covered by an h-BN protective layer. The devices are annealed in vacuum for 1 hour at 300°C. Figure 8(b) & (c) shows the device morphology change before and after the annealing. It is found that the exposed black P devices almost completely disappear, and the h-BN-covered devices are also partially removed. In contrast, the h-BN protective layer remains intact after the annealing. Subsequent electronic measurements show no conductivity between any electrode pairs. Therefore, annealing temperatures appropriate for other 2D materials (*e.g.*, $MoS_2$) may not apply to black P devices.

We further explore annealing black P device at lower temperature. Figure 8 (d)-(f) show the electronic performance of another black P FET (not covered by h-BN) made using the same dry transfer technique after subsequent 200°C vacuum annealing. Under optical microscope the device appearance remains unaltered throughout the process, while its electronic performance goes through a series of changes: both its conductivity and gate tunability increases after the first two annealing sessions, and then becomes non-conductive after additional annealing (note the scale on the vertical axis, which is zoomed in). Our results show that the morphology changes (via optical microscope inspections) are not the only indications of device degradation; the electronic performance can alter despite the device appearance in the optical microscope remaining unchanged.

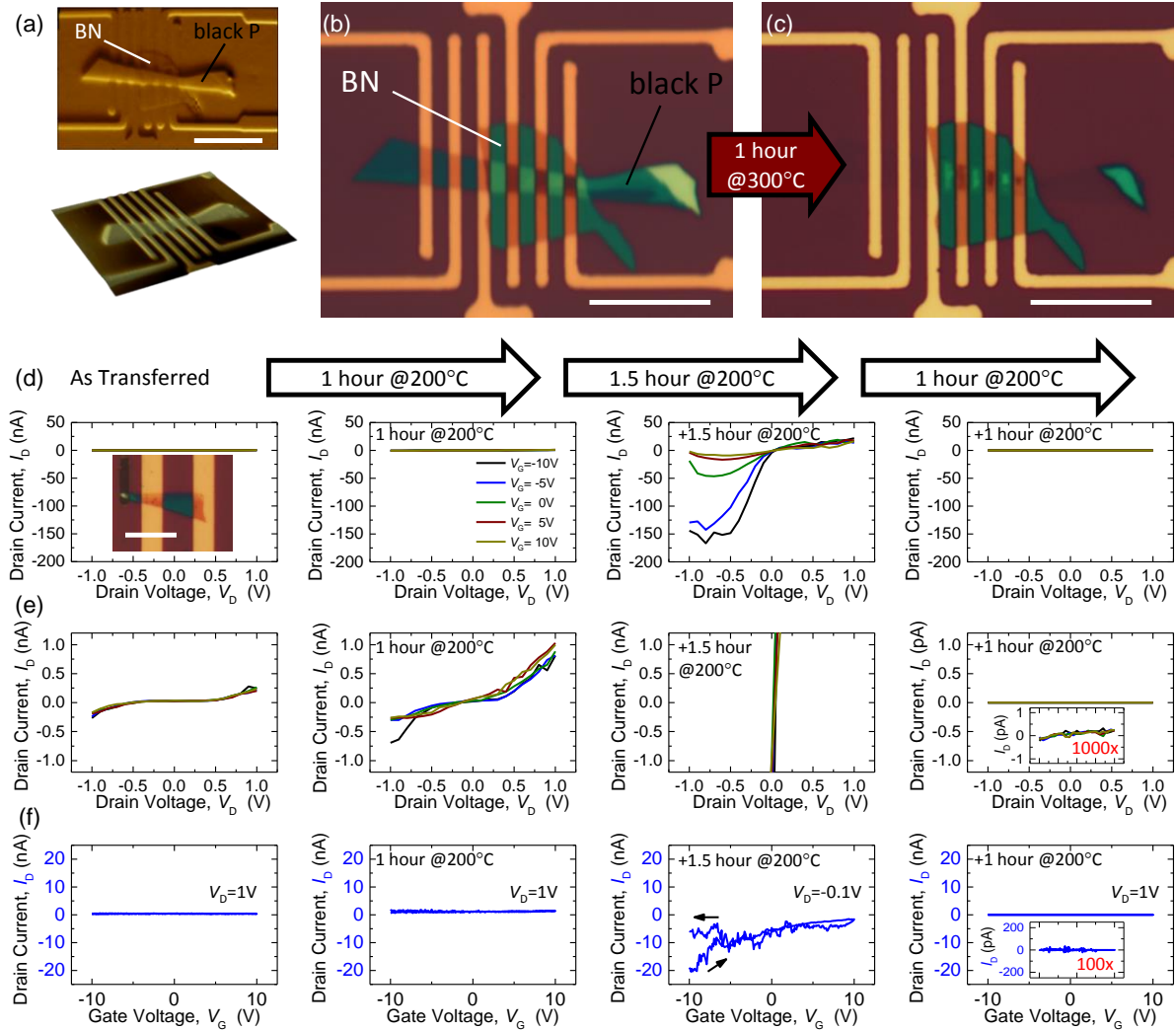

**Figure 8.** Effect of thermal annealing on black P FET transistors. (a) AFM images of a group of black P FETs. Several of the devices are covered by an h-BN protective layer. (b) Optical image before and (c) after 300°C vacuum annealing for the same device in (a). (d) Evolution of $I_D$ vs. $V_D$ for an uncovered black P FET device under a series of 200°C vacuum annealing sessions. The inset in the first pane shows the optical image of the device. Legends in the second pane from left apply to all panes in (d) and (e). (e) Same as (d), with the vertical axis rescaled to show details at small currents. Note the inset of the rightmost pane is further zoomed in the vertical direction. (f) $I_D$ vs. $V_G$ for the same device in (d). The $V_D$ values used in measurements are labeled in each pane. Note the inset of the rightmost pane is further zoomed in the vertical direction.

## E. *Observation of Electrical Alteration of Device Morphology*

Besides environmental variables and thermal annealing treatment, we have also studied device performance upon high electrical gating. Figure 9 shows the modification in device morphology of a suspended-channel black P FET device with multiple electrodes. By sweeping the gate polarization voltage between $V_G$ = -30V and +30V while limiting the source-drain bias to $I_D$=1µA through a pair of opposing electrodes (indicated by red dots in Fig. 9(a)), the device exhibits significant changes in appearance (Fig. 9(a)(c)(e) vs. Fig. 9(b)(d)(f)): the non-suspended portion of the black P flake becomes highly corrugated, and the suspended portion largely breaks off along its circumference.

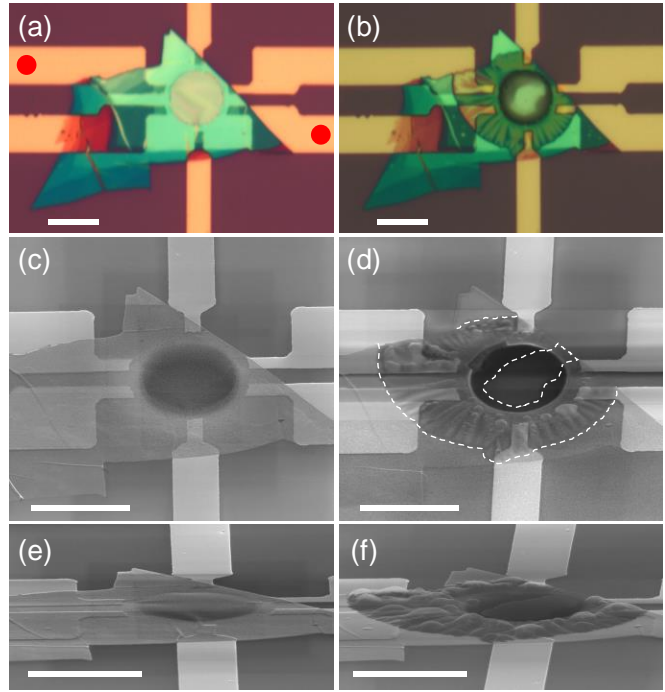

**Figure 9.** Electrical alteration of black P FET transistor. Shown are the (a), (b) optical, (c), (d) aerial view SEM, (e), (f) tilted side view SEM images of a ~20nm-thin suspended black P FET device before and after 1µA current passing through a pair of electrodes (red dots in (a)). Dashed lines in (d) show the boundary of the corrugated part and suspended section after the treatment. Scale Bars: 10µm.

The center region of the device is originally suspended over a 10µm-wide, 2.2µm-deep circular microtrench. We perform finite elemental modeling (FEM, using COMSOL) simulation for this device structure, and found maximum deflection ~50nm and maximum strain ~$10^{-4}$ at the center of the black P diaphragm under 30V gate bias. It is therefore unlikely that electrostatic pull-in could take place. In contrast, we observe transient current through the gate electrode at high gate voltage (>15V), suggesting that electrical leakage from the gate may be responsible for the observed effect. While in theory the $SiO_2$ layer should withstand much higher gate bias, in practice the 2.2µm-deep microtrench is fabricated by dry etching away the 290nm $SiO_2$ followed by dry etching 1.9µm into the Si substrate. During the Si dry etch process, sulfur hexafluoride ($SF_6$) is ionized and radicalized by the plasma and creates a mixture of $SF_x$ and $F_y$ ions and neutrals, where $x$ and $y$ can range from 0 to 6 and 1 to 2, respectively.[26] The process is a combination of physical (anisotropic) and chemical (isotropic) etching. The physical process mostly sputters materials off the bottom of the trench, and could lead to material redeposition on the sidewall. Since the Si dry etch takes place after the sidewall of $SiO_2$ is formed, it is possible that during this process some Si material is redeposited onto the $SiO_2$ sidewall. Such Si impurities on the $SiO_2$ sidewall, though buried by the subsequent passivation layer, could lead to degradation in the electrical breakdown voltage of the $SiO_2$ layer by forming discrete conductive paths.

By applying a large gate bias (up to 30V), a breakdown current can start to flow vertically along the sidewall of the microtrench (from the Si back gate to the suspended black P crystal). This is consistent with the observation (Fig. 9) that the morphology change in the crystal is centralized around the microtrench area, largely breaking off the suspended region.

Surprisingly, while the device appearance after the high gating process shows significant morphology change and may suggest catastrophic device degradation, the electronic performance of the black FET remains not only measurable, but it also exhibits clear P-type semiconductor behavior with reasonable conductance (Fig. 10), across several different pairs of electrodes (note that during these measurements, the gate voltage is limited to within ±10V, and no leakage current through the gate is observed). Once again, the device appearance does not directly tell the device performance. Note that such observation may not be accounted for entirely by the residue conductance from the peripheral area whose appearance remain unchanged, as some of the measurements are made across electrode pairs which are only connected via area with altered morphology. While our experiment leads to clear, interesting observations, additional characterization and analysis are required to understand the nature of the resulting device and the observed electrical alteration in black P FET devices.

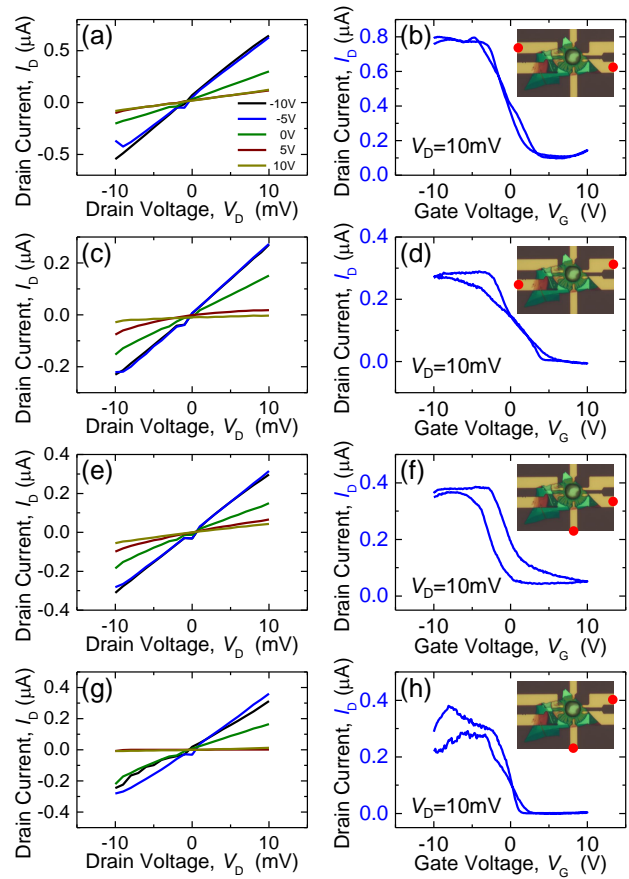

**Figure 10.** Performance of electrical alteration of black P FET transistor. (a)-(h) The $I_D$ vs. $V_D$ and $I_D$ vs. $V_G$ characteristics of the electrically altered device. Measurements are made across different electrode pairs as indicated by the red dots in the insets. The $V_G$ values for all the $I_D$ vs. $V_D$ measurements are indicated by the legend in (a).

## III. Conclusions

In conclusion, we have performed month-long detailed electronic characterization of black P FETs under a number of environmental configurations. The degradation in device performance precedes device appearance change, and can occur in ambient even in the absence of light illumination. Black P devices with h-BN protective layer covering the top surface are not immune to, but show deferred, degradation through exposure to environment. Both thermal annealing and passing electrical current can alter the devices performance and appearance. Our results suggest that black P electronic devices are susceptible to a number of environmental variables, including established procedures compatible with other 2D devices. Therefore, improved device designs (such as fully-encapsulated geometry) and new procedures in fabrication, storage, and measurement are required toward stable operation of black P electronic devices.

**Acknowledgement**: We thank the support from Case School of Engineering, National Academy of Engineering (NAE) Grainger Foundation Frontier of Engineering (FOE) Award (FOE2013-005), National Science Foundation CAREER Award (ECCS-1454570), CWRU Provost's ACES+ Advance Opportunity Award. Part of the device fabrication was performed at the Cornell NanoScale Science and Technology Facility (CNF), a member of the National Nanotechnology Infrastructure Network (NNIN), supported by the National Science Foundation (Grant ECCS-0335765).

Note: While we were processing the extensive data and preparing the manuscript, a few related studies became available.[19,27,28,29] The observations are in general agreement with our results and prediction that alternative geometries such as encapsulation can lead to better protection. However, in the present systematic study, for the first time, we have showed detailed time evolution of the electronic performance, examined the effects from individual ambient components (air, light, humidity), and investigated the effects and degradation due to thermal annealing and passing electrical current.